\documentstyle[12pt,twoside,fleqn,espcrc1,epsfig]{article}

\newcommand{\AmS}{{\protect\the\textfont2
  A\kern-.1667em\lower.5ex\hbox{M}\kern-.125emS}}

\title{Chiral dynamics of the nuclear equation of state}
\author{Matthias Lutz \address{GSI, 64220 Darmstadt, Germany }}

\begin{document}
% typeset front matter
\maketitle

\begin{abstract}
We present a new chiral power expansion scheme for the nuclear 
equation of state. The scheme is effective in the sense that it is 
constructed to work around nuclear saturation density. The leading  
and subleading terms are evaluated and are shown to provide an 
excellent equation of state. As a further application we considered 
the chiral quark condensate in nuclear matter. Already at nuclear 
saturation density we predict a substantially smaller reduction of 
the condensate as compared to conventional approaches. 
\end{abstract}

\section{Introduction}

The nuclear equation of state lies at the heart of nuclear physics. 
It is therefore of considerable importance to find the appropriate 
form of chiral perturbation theory ($\chi $PT), the most powerful 
tool of modern nuclear physics, and apply it systematically to the 
nuclear many body problem. 

A key element of any microscopic theory for the nuclear equation of 
state is the elementary nucleon nucleon scattering process. In the 
context of $\chi $PT  this problem was first addressed by Weinberg 
who proposed to derive a chiral nucleon nucleon potential in time 
ordered perturbation theory \cite{Weinberg}. The NN-phase shifts 
then follow from the solution of the Schroedinger equation 
appropriately fed with the chiral potential. This program was 
carried out by Ordonez, Ray and  van Kolck \cite{Bira}. Weinberg's 
scheme is plagued by two problems. First the systematic 
renormalization of the chiral potential scheme is an open problem 
and not resolved. The scheme is regularization scheme dependent. 
Second the use of a potential approach in nucleon nucleon 
scattering can be justified only if retardation effects are small. 

To overcome both problems the present author proposed first to 
apply chiral power counting rules to the nucleon nucleon scattering 
amplitude directly \cite{Lutz}. This approach cures both problems. 
First the manifest covariant version of the chiral Lagrangian can 
be applied. A relativistic form of $\chi $PT was applied before by 
Gasser \cite{Gasser} in the 1-nucleon sector but rejected and 
replaced by the heavy mass formulation of $\chi $PT \cite{Manohar}. 
There are two problems inherent with the relativistic approach of 
\cite{Gasser}. First, any covariant derivative acting on the 
nucleon field produces the large nucleon mass, $m_N$, and therefore 
must be assigned the minimal chiral power zero. Thus an infinite 
tower of interaction terms need to be evaluated at a given finite 
chiral order. Second, the straightforward evaluation of 
relativistic diagrams involving nucleon propagators generates 
positive powers of the nucleon mass from loop momenta larger than 
the nucleon mass. The chiral power counting rules are spoiled. 

The two problems are solved as follows. Rather than performing the 
$1/m_N$ expansion at the level of the Lagrangian density, as 
achieved in the heavy mass formulation of $\chi $PT, it is equally 
well possible to work out this expansion explicitly at the level of 
any individual relativistic Feynman diagrams \cite{Lutz}. The 
second problem is overcome by a suitable reorganization of the 
infinite tower of chiral interaction terms. Of course we expect our 
relativistic scheme to be equivalent to the heavy fermion 
formulation of chiral perturbation theory in the one-nucleon 
sector. 

The second important observation made first in \cite{Lutz} is the 
fact that the chiral power counting rules can be generalized for 
2-nucleon reducible diagrams. For a given diagram each pair of 
intermediate nucleons causes a reduction of one chiral power as 
compared to the 'naive' chiral power. The non perturbative 
structures like the deuteron bound state are generated naturally  
since the bare 2-nucleon vertex is renormalized strongly so that it 
effectively carries chiral power minus one. 

An attempt to apply the generalized chiral power counting rules of 
\cite{Lutz} to the nuclear many body problem quickly reveals that 
even though the pion dynamics remains perturbative the local 
2-nucleon interaction requires extensive resummations. 

\section{Chiral expansion scheme for nuclear matter}

It is straightforward to combine the chiral expansion of the 
nucleon scattering amplitudes with the low density expansion. Naive 
application to the energy per particle, $\bar E(k_F)$,  of isospin 
symmetric nuclear matter would lead to an expansion of the form:
\begin{eqnarray}
\bar E(k_F) &=& \sum_n\, \bar E_n \!\!\left(\frac{k_F}{m_\pi}
\right)  \left( \frac{k_F}{\Lambda }\right)^n
\label{exp1}
\end{eqnarray}
with $\Lambda \neq m_\pi $ and $\rho=2\,k_F^3/(3\,\pi^2)$. The 
expansion coefficients $\bar E_n$ are functions of the ratio 
$k_F/m_\pi $. Such a scheme is obviously restricted to extremely 
low  density, if useful at all, since the typical scale $\Lambda $ 
is small. This follows for example if one considers the 2-nucleon 
rescattering contribution to the 3-nucleon scattering process. Then 
the deuteron pole term induces structures like 
$k_F/\sqrt{m_N\,\epsilon_D }$ where $\epsilon_D 
\simeq 2 $ MeV denotes the deuteron binding energy. One is lead to 
identify $\Lambda \sim \sqrt{m_N\,\epsilon_D }$. 

A scheme working at nuclear saturation density therefore requires a 
further resummation. Obviously an expansion of the form 
\begin{eqnarray}
\bar E(k_F) &=& \sum_n\, \bar E_n\!\!\left(\frac{k_F}{m_\pi}, 
\frac{k_F}{\Lambda_S}\right)  \left( \frac{k_F}{\Lambda_L}\right)^n
\label{exp2}
\end{eqnarray}
with typical small scales, $\Lambda_S$, like  $ \sqrt{m_N 
\,\epsilon_D} $  and typical large scales 
$\Lambda_L \simeq 4\,\pi \,f_\pi \simeq 1$ GeV must be achieved. 
The expansion coefficients $\bar E_n$ are complicated and hitherto 
unknown functions of the Fermi momentum $k_F$. They can be computed 
in terms of the free space chiral Lagrangian furnished with a 
systematic resummation technique \cite{Lutz:prep}. 

In this work we present a somewhat less microscopic approach in 
spirit close to the Brueckner scheme but more systematic in the 
sense of effective field theory. Since  the typical small scale 
$\Lambda_S $ is much smaller than the Fermi momentum $\bar k_F 
\simeq 265 $ MeV at nuclear saturation density, one may expand the 
coefficient functions $\bar E_n$ around $\bar k_F $ in the 
following manner 
\begin{eqnarray}
\bar E_n\!\!\left(\frac{k_F}{m_\pi}, 
\frac{k_F}{\Lambda_S}\right) &=& \bar E_n\!\!\left(\frac{k_F}{m_\pi}, 
\frac{\bar k_F}{\Lambda_S}\right)
+\sum_{k=1}^\infty \bar E_n^{(k)}\!\!\left(\frac{k_F}{m_\pi} 
\right) 
\left( \frac{\Lambda_S }{k_F}-\frac{\Lambda_S}{\bar k_F}\right)^k
\; .
\label{exp3}
\end{eqnarray}
Note that we do not expand in the ratio $m_\pi/k_F $. If one 
expanded also in this ratio $m_\pi/k_F $ one would arrive at the 
Skyrme phenomenology \cite{Skyrme} applied successfully to nuclear 
physics many years ago. It should be clear that this scheme is 
constructed to work around nuclear saturation density but will fail 
at small density. We note that also conventional approaches like 
the Walecka mean field or the Brueckner scheme are known to be 
incorrect at small density. 

Technically our scheme can be generated by the effective Lagrangian 
density 
\begin{eqnarray}
{\mathcal L}_{int}(k_F) &=& \frac{g_A}{2\,f_\pi} \, \bar N
\gamma_5\, \gamma^\mu \cdot 
\left( \partial_\mu  \,{\vec \pi }\right) \cdot {\vec \tau} \,N
\nonumber\\
&+& \frac{1}{8\,f_\pi^2}\,\left(g_0(k_F)+\frac{1}{4}\,g_A^2 \right)
\left( \bar N \,\gamma_5\,\tau_2\, C^{-1}\,\bar N^t\right)
\left( N^t\, C\,\tau_2\,\gamma_5\,N \right)
\nonumber\\
&+& \frac{1}{8\,f_\pi^2}\,\left(g_1(k_F)+\frac{1}{4}\,g_A^2 \right)
\left( \bar N \,\gamma_\mu\,\vec \tau \,\tau_2\, C^{-1}\,\bar N^t\right)
\left( N^t\, C\,\tau_2\,\vec \tau \,\gamma^\mu\,N \right)
\label{l1}
\end{eqnarray}
where the couplings $g_0=g_0(k_F)$ and $g_1=g_1(k_F)$ are density 
dependent. A more systematic derivation of the expansion 
(\ref{exp2}) and (\ref{exp3}) applying suitable resummation 
techniques will be presented elsewhere \cite{Lutz:prep}. 

\begin{figure}[b]
\epsfysize=1.5cm
\begin{center}
\mbox{\epsfbox{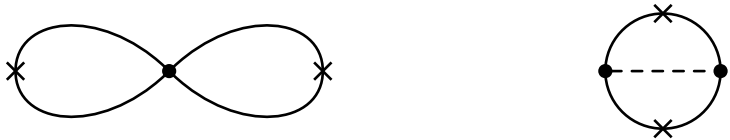}}
\end{center}
\caption{Leading contribution of chiral order $Q^6$.}
\label{fig2}
\end{figure}

The chiral power counting rules are simplified significantly as 
compared to a fully microscopic scheme. The presence of a further 
small scale $k_F \sim Q \sim m_\pi $ does not anymore generate an 
infinite tower of diagrams, to be considered at a given chiral 
order, since by construction the troublesome local 2-nucleon vertex 
must not be iterated, i.e. terms proportional to $g_0^n(k_F)$ and 
$g_1^n(k_F)$ with $n>1$ are already included in $g_0(k_F)$ and 
$g_1(k_F)$ and therefore must not be considered. The pion dynamics, 
if properly renormalized, remains perturbative like in the vacuum 
case. Consider for example the two loop diagrams depicted in Fig. 
\ref{fig2} 
where the nucleon line with a 'cross' represents the projector onto 
the Fermi sphere 
\begin{eqnarray}
\Delta S_N(p) = \left( \gamma \cdot p+m_N \right)
2\,\pi\,i\,\Theta(p_0)\,\delta (p^2-m_N^2)\,
\Theta \left(k_F^2-\vec p\,^2\right)\; .
\label{cross}
\end{eqnarray}
The first diagram in Fig. \ref{fig2} is proportional to 
$g_{0,1}(k_F) \, k_F^6$ and is therefore ascribed the chiral order 
$Q^6 $ since the effective vertex $g_{0,1}(k_F) \sim Q^0 $ carries 
chiral power zero. The second diagram, the one pion exchange 
contribution, is also of chiral order $Q^6$ since it is 
proportional to $k_F^6 $ multiplied with some dimensionless 
function $f(m_\pi/k_F)$. 
\begin{figure}[t]
\epsfysize=5.5cm
\begin{center}
\mbox{\epsfbox{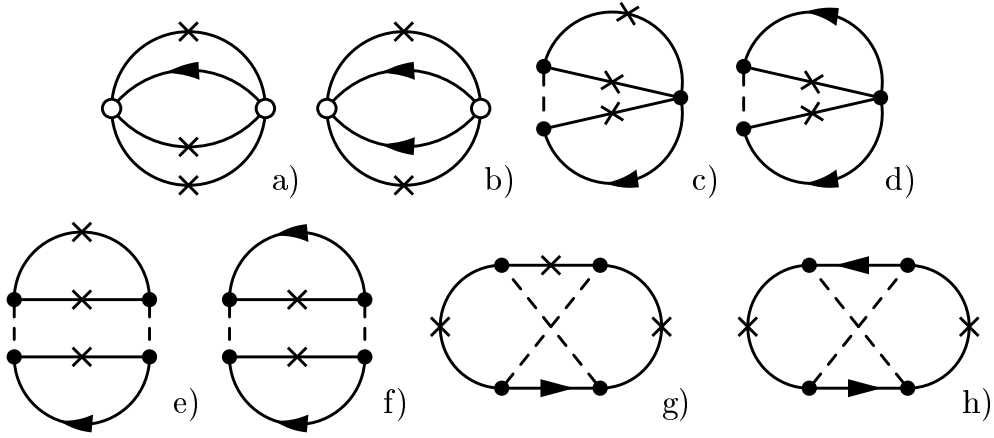}}
\end{center}
\caption{Subleading contribution of chiral order $Q^7$.}
\label{fig3}
\end{figure}
In Fig. \ref{fig3} we collected all diagrams of chiral order $Q^7$. 
Here we introduced two types of 2-nucleon vertices. The filled 
circle represents the full vertex of (\ref{l1}) proportional  to 
$g_{0,1}+g_A^2/4 $ and the open circle the counter term 
proportional to $g_A^2/4 $. The dashed line is the pion propagator 
and the directed solid line the free space nucleon propagator. We 
point out that the diagrams b), d), f) and h) in Fig. \ref{fig3} 
are divergent. The leading chiral contribution of the sum of all 
diagrams, however, is finite. 
\begin{figure}[t]
\epsfysize=8cm
\begin{center}
\mbox{\epsfbox{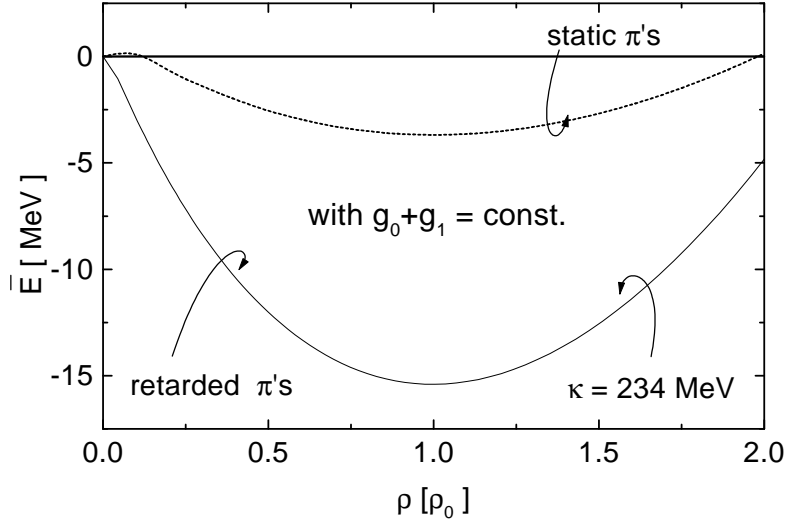}}
\end{center}
\caption{The equation of state for isospin symmetric nuclear matter.}
\label{fig5}
\end{figure}

It is instructive to also display the non relativistic form of the 
Lagrangian density (\ref{l1}) as suggested by Weinberg 
\cite{Weinberg} in the context of the nucleon nucleon scattering 
problem  
\begin{eqnarray}
{\mathcal L}_{int}(k_F) &=&\frac{g_A}{2\,f_\pi} \, \bar N
\left(\vec \sigma \cdot \vec \nabla \right)\Big(
{\vec \pi } \cdot {\vec \tau} \Big)\,N
\nonumber\\
&+& \frac{1}{8\,f_\pi^2}\,\left(g_0(k_F)+\frac{1}{4}\,g_A^2 \right)
\left( \bar N \,\vec \sigma\,\sigma_2\,\tau_2\,\bar N^t\right)
\left( N^t\,\tau_2\,\sigma_2\,\vec \sigma \,N \right)
\nonumber\\
&+& \frac{1}{8\,f_\pi^2}\,\left(g_1(k_F)+\frac{1}{4}\,g_A^2 \right)
\left( \bar N \,\sigma_2\,\,\vec \tau \,\tau_2\,\bar N^t\right)
\left( N^t\, \tau_2\,\vec \tau \,\sigma_2\,N \right)
\label{l2}
\end{eqnarray}
with $N$ now a two component spinor field and $\bar N =N^\dagger $ 
for notational convenience. According to Weinberg the two particle 
reducible diagrams are to be evaluated with static pions. We 
therefore evaluate all diagrams of Fig. \ref{fig3} with the 
interaction vertices of (\ref{l2}) and static pion propagators. The 
solid line with a 'cross' now represents the non relativistic limit 
of (\ref{cross}) and the directed solid line the free non 
relativistic nucleon propagator.  We remind the reader that in the 
relativistic approach (\ref{l1}) we perform the chiral expansion at 
hand of a given Feynman diagram. For technical details we refer to 
\cite{Lutz:prep}. Both schemes indeed lead to identical results for 
all diagrams except diagram h). Only in the extreme low density 
limit we find agreement of the non relativistic with the  
relativistic scheme also for diagram h). We conclude that it is 
incorrect to use static pions in two-particle reducible diagrams. 
Therefore the use of a potential in the $\chi $PT  approach of the 
nuclear many body problem {\it or} the nucleon nucleon scattering 
problem cannot be justified. A relativistic approach is required. 

In Fig. \ref{fig5} our result for the isospin symmetric nuclear 
equation of state is shown.  We compare our relativistic approach 
(\ref{l1}) with the static potential approach of (\ref{l2}). The 
relevant coupling $g_0+g_1$ is adjusted to obtain nuclear 
saturation at $k_F = 265 $ MeV. The non relativistic scheme gives a 
rather poor equation of state as shown in Fig. \ref{fig5}. The 
binding energy and the incompressibility are too small. We 
emphasize that the coupling functions $g_{0,1}(k_F)$  are to be 
determined from the nuclear equation of state. It is therefore 
legitimate to cure the 'non relativistic' equation of state by  
giving the coupling  a residual density dependence 
\begin{eqnarray}
g_0(k_F) +g_1 (k_F)&=& \gamma_0+ \gamma_1\,\frac{\bar k_F}{k_F}
\left(1-\frac{k_F}{\bar k_F} \right)+
\gamma_2\,\left(\frac{\bar k_F}{k_F}\right)^2
\left(1-\frac{k_F}{\bar k_F} \right)^2 + \cdots\,.
\label{}
\end{eqnarray}
However there is a strong consistency constraint: according to our 
scale argument (\ref{exp3}) the density dependence of the couplings 
$g_{0,1}(k_F)$ must be weak. If the nuclear saturation required a 
strong density dependence our scheme had to be rejected. We turn to 
the relativistic scheme. Here the set of parameters $g_0+g_1 
\simeq 2.7 $, $g_A \simeq 1.2 $, $m_\pi \simeq 140 $ MeV  and 
$f_\pi \simeq 93 $ MeV give an excellent result for the equation of 
state. The empirical saturation density $\bar k_F \simeq 265 $ MeV 
and the empirical binding energy of 16 MeV are reproduced. The 
incompressibility with $\kappa \simeq 234 $ MeV is also compatible 
with the empirical value $(210 \pm 30) $ MeV of \cite{Blaizot}. We 
conclude that in fact the consistency constraint points towards the 
correct scheme (\ref{l1}). 

\section{The chiral order parameter $\langle \bar qq \rangle $}

The quark condensate, $\langle \bar qq \rangle (\rho )$, is an 
object of utmost interest. It measures the degree of chiral 
symmetry restoration in nuclear matter. Furthermore it is an 
important input for QCD sum rules \cite{S.H.Lee} or the Brown Rho 
scaling hypothesis \cite{BR-scaling}. According to the Feynman 
Hellman theorem the quark condensate can be extracted unambiguously 
from the total energy, $E(\rho) $, of nuclear matter once the 
current quark mass dependence of $E(\rho , m_Q )$ is known 
\begin{eqnarray}
\langle \bar qq \rangle (\rho ) &=& \frac{1}{V}\,
\frac{\partial}{\partial\,m_Q} \, E(\rho, m_Q )
\label{FH1}
\end{eqnarray}
with $m_Q=m_u=m_d$. Recall that $ E(\rho )/V 
= ( m_N+\bar E(\rho ) )\,\rho $ is determined by the nuclear 
equation of state $\bar E(\rho ) $. Since our chiral approach was 
set up to treat the pion dynamics and therewith the current quark 
mass dependence of the equation of state properly it is very much 
tailored to be applied to the quark condensate.  

It is convenient to consider the relative change of the quark 
condensate since it is  renormalization group invariant: 
\begin{eqnarray}
\frac{\langle \bar q\, q \rangle (\rho)}
{\langle \bar q\, q \rangle (0)}
=1-\frac{\Sigma_{\pi N}\,\rho}{m_\pi^2\,f_\pi^2}
-\frac{\alpha_\pi(\rho)\,\rho}{2\,m_\pi\,f_\pi^2}
\label{FH2} \; .
\end{eqnarray}
The second term in (\ref{FH2}) follows from the nucleon rest mass 
contribution to the total energy of nuclear matter together with 
the definition of the pion nucleon sigma term, $\Sigma_{\pi N} $: 
\begin{eqnarray}
\Sigma_{\pi N} = m_Q \,\langle N|\bar q\, q |N\rangle
=m_Q\,\frac{d\,m_N}{d\,m_Q} \; .
\label{}
\end{eqnarray}
The last term in (\ref{FH2}) measures the sensitivity of the 
nuclear equation of state, $\bar E(\rho, m_\pi )$, on the pion mass 
\begin{eqnarray}
\alpha_\pi(\rho) &=& 
-\frac{2\,m_\pi\,f_\pi^2}{\langle \bar q\, q \rangle (0)}
\,\frac{\partial }{\partial\, m_Q}\,\bar E(\rho ,m_\pi )
=\left(1+{\mathcal 
O}\left( m_\pi^2 \right) \right)\frac{\partial }{\partial 
\,m_\pi}\,\bar E(\rho, m_\pi ) 
\label{}
\end{eqnarray}
where we consider now the current quark mass $m_Q=m_Q(m_\pi)$ as a 
function of the physical pion mass. We emphasize that the second 
term in (\ref{FH2}) written down first in 
\cite{Drukarev,Lutz92,Cohen} does not probe the nuclear many body 
problem and therefore should be considered with great caution. It 
is far from obvious that this term is the most important one at 
nuclear saturation density. The only term susceptible to nuclear 
dynamics is the last term in (\ref{FH2}). 

\begin{figure}[t]
\epsfysize=8cm
\begin{center}
\mbox{\epsfbox{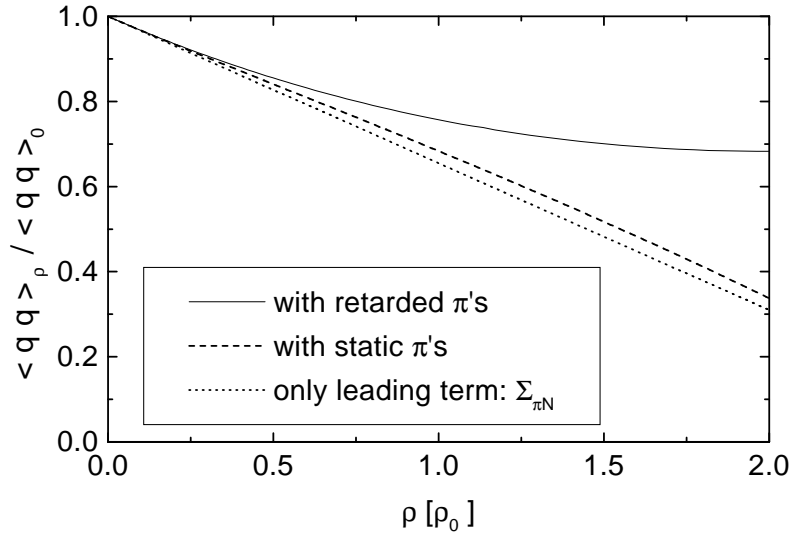}}
\end{center}
\caption{The quark condensate in isospin symmetric nuclear matter.}
\label{fig6}
\end{figure}

In Fig. \ref{fig6} we present our result for the quark condensate 
in nuclear matter. We confront the 'leading' term driven by 
$\Sigma_{\pi N}\simeq 45 $ MeV with first, our result applying 
static pions and second, our result including important retardation 
effects. The former result is rather close to the 'leading' order 
result. The nuclear many body dynamics as induced by static pions  
appear to effect the quark condensate little. This confirms results 
obtained in the Brueckner \cite{Li} and Dirac-Brueckner 
\cite{Brockmann} approach. Note that neither the Brueckner nor the 
the Dirac-Brueckner approach includes pionic retardation effects 
systematically. We celebrate the central result of this section: as 
shown in Fig. \ref{fig6} the inclusion of pionic retardation 
effects leads to a significantly less reduced quark condensate in 
nuclear matter. 

Our result implies strong consequences for the QCD sum rule 
approach to the properties of hadrons in nuclear matter. 

\section*{Acknowledgments}

The author acknowledges collaboration with Christoph Appel and 
Bengt Friman on part of this work. A joint publication, where the 
present scheme is applied also to the  nuclear Landau parameters, 
is in preparation.

\end{document}